\documentclass[prx,onecolumn,english,floatfix,nofootinbib]{revtex4-2}

\usepackage[utf8]{inputenc}
\setcitestyle{authoryear,round, sort}
\usepackage{enumitem} 
\usepackage{textgreek} 
\usepackage{physics} 
\usepackage{amsmath}
\usepackage{amsthm} 
\usepackage{amssymb}
\usepackage{mathtools} 
\usepackage[scr=boondoxo]{mathalfa}
\usepackage{pst-node}%
\DeclareMathAlphabet{\mathpzc}{OT1}{pzc}{m}{it}

\newcommand{\p}{\mathbb{P}}
\newcommand{\m}{\mathscr{p}_w}

\newcommand{\w}{\mathcal{P}}
\newcommand{\g}{\mathcal{G}}
\newcommand{\s}{\tilde{\mathscr{p}}_w}

\newcommand{\x}{\mathscr{p}}

\usepackage{parskip}

\usepackage[left=2cm, right=2cm, top=2.1cm, bottom=2.1cm]{geometry} 
\usepackage{graphicx} 
\usepackage{grffile}
\graphicspath{{Figures/}} 
\usepackage{multirow} 

\usepackage{siunitx} 

\usepackage{hyperref} 
\hypersetup{
    colorlinks=true,
    linkcolor=violet,
    filecolor=violet,      
    urlcolor=violet,
    citecolor=violet
}

\usepackage{adjustbox}
\newcolumntype{?}{!{\vrule width 1.5pt}}

\usepackage{anyfontsize}

\DeclareMathSizes{10}{10}{10}{10}

\usepackage{mathtools}
\date{\today}
\usepackage{xcolor}

\newcommand\summaryname{ABSTRACT}
\newenvironment{Abstract}%
    {\small\begin{center}%
    \bfseries{\summaryname} \end{center}}

\begin{document}

\title{How weak values illuminate the role of “hidden”-variables as predictive tools}

\author{Xabier Oianguren-Asua}
\affiliation{Department of Physics, Universitat Aut\`onoma de Barcelona, 08193 Bellaterra, Spain}%
\author{Albert Solé\footnote{The research conducted by AS is funded by Grant PID$2020$-$115114$GB-I$00$ and Grant CEX$2021$-$001169$-M from  MCIN$/$AEI$/10.13039/501100011033$.}}
\affiliation{Serra Húnter Fellow, Departament de Filosofia, Universitat de Barcelona, Spain; BIAP}
\author{Carlos F. Destefani\footnote{The research conducted by CD and XO is funded by Grant No. PID$2021$-$127840$NB-I$00$ (MCIN$/$AEI$/$FEDER, European Union).}}
\affiliation{Department of Electronic Engineering, Universitat Aut\`onoma de Barcelona, 08193 Bellaterra, Spain}%
\author{Xavier Oriols}
\email{xavier.oriols@uab.cat}
\affiliation{Department of Electronic Engineering, Universitat Aut\`onoma de Barcelona, 08193 Bellaterra, Spain}%

\maketitle
\vspace{-0.2cm}
\begin{Abstract}

In this chapter we offer an introduction to weak values from a three-fold perspective: first, outlining the protocols that enable their experimental determination; next, deriving their correlates in the quantum formalism and, finally, discussing their ontological significance according to different quantum theories or interpretations. We argue that weak values have predictive power and provide novel ways to characterise quantum systems. We show that this holds true regardless of ongoing ontological disputes. And, still, we contend that certain “hidden” variables theories like Bohmian mechanics constitute very valuable heuristic tools for identifying informative weak values or functions thereof. To illustrate these points, we present a case study concerning quantum thermalization. We show that certain weak values, singled out by Bohmian mechanics as physically relevant, play a crucial role in elucidating the thermalization time of certain systems, whereas standard expectation values are “blind” to the onset of thermalization.

\end{Abstract}
\vspace{0.1cm}
 {\small {\bf Keywords:} Weak values, local expectations, Bohmian mechanics, laboratory protocol, microscopic characterization.}
 \vspace{0.3cm}

\parskip=0.2cm
{\small \tableofcontents}
\setlength{\parskip}{0.3 cm}

\newpage
\section{Introduction}\vspace{-0.1cm}

\setcounter{page}{1}

In the literature, {\em weak values} \citep{original, eliahu} have traditionally been discussed either mostly from an ``ontological perspective" \citep{fankhauser}, or from an ``experimental perspective" \citep{desgracia1, desgracia2}. The lack of a clear distinction between the experimental, the ontological and the formal descriptions of weak values has led many physicists to suspect that weak values are either mere theoretical oddities with no practical import or experimental ``tricks" lacking serious theoretical significance. In this chapter, we provide a comprehensive description of weak values that is suitable for physicists with different interests, ranging from those focused on their operational aspects to those interested in their ontological implications.\footnote{We refer to a physicist who demands a fully-fledged description of reality from quantum mechanics as having an “\textit{ontological} attitude". In contrast, we say that a physicist who uses quantum mechanics merely as a predictive algorithm has an “\textit{empiricist }attitude”.} To achieve this, we explicitly distinguish (i) their {\em experimental} description in terms of laboratory protocols and statistical post-processing, (ii) their {\em formal} description within the general state-operator formalism, and (iii) their {\em ontological} significance according to different quantum theories or interpretations. 

By considering the first two descriptions, we draw the first main conclusion of this chapter: weak values and their functions, along with the associated operational techniques, enable numerous new predictions and provide experimental access to a wealth of information about microscopic systems. Typically, this information is not attainable with more conventional characterization tools like expectation values of usual operators. Through an insightful case study, we demonstrate the remarkable characterization ability of weak values, even in situations where conventional expectation values fail to elucidate the relevant properties of a quantum system.

Moreover, this same case study serves as an illustration of the second main conclusion of the chapter: the predictive power of weak values remains unaffected by debates on their ontological status. We should remind that such an emphasis on practicality is not novel. For instance, when dealing with the {\em wavefunction}, it has already been widely accepted and has proven to be highly productive, as witnessed by the impressive developments in quantum technologies regardless of ongoing debates on the meaning of the wavefunction. We consider that similar developments can occur if the usefulness of weak values as predictive and characterisation tools is duly recognized.

This does not entail, however, that ontological discussions are completely idle in this context. Apart from their obvious interest in the foundations of quantum mechanics, we contend that certain quantum theories or interpretations, even if motivated through ontological arguments, can be of interest for those with an ``empiricist attitude". There are certain weak values or functions thereof that are singled out as physically relevant by Bohmian mechanics \citep{bohm, Holland, JordiXavier, Durr} and other “hidden-variable” theories, whereas they do not hold any particular significance from the standpoint of other approaches. We show by means of the case study that the specific weak values highlighted by some of these “hidden” variable theories are valuable for characterizing problematic quantum systems, a characterization that can be contrasted in the laboratory. This illustrates that such theories provide heuristically convenient frameworks to analyse quantum systems both theoretically and experimentally. As a consequence, we draw the third main conclusion of the chapter: one may use certain “hidden” variable theories, such as Bohmian mechanics, as heuristic tools in the experimental characterization of quantum systems, even if one is not necessarily committed to the ontology of such theories.

\section{On the meaning of weak values}

As stated in the Introduction, in this section we aim to characterize three defining facets of weak values, namely, their {\em experimental}, {\em formal} and {\em ontological} descriptions. In order to clarify this three-fold distinction, it is useful to apply it first to  standard {\em expectation values} which are better understood than weak values. Regarding the {\em experimental} description\footnote{In the philosophical literature, an operational reduction of a concept is sometimes understood as the {\em definition} of that concept in purely {\em observational} terms or by specifying a set of operations in a laboratory. We are {\em not} attempting to provide such an analysis here. First, the operations involved in the protocols we describe do not solely concern purely observational quantities or methods. They require a significant amount of theory, including bridge principles of the predictive core of quantum mechanics, such as those related with the preparation of systems. Second, in our experimental description of both standard expectations and weak values we resort to notions, such as ``laboratory measurements", that, for lack of space, we cannot fully analyse in terms of specific laboratory operations. Now, even if the experimental description of weak values is theory-laden, the amount of theory involved is much lower than that required for the discussion of the formal and ontological aspects of weak values.} of expectation values (choosing the momentum's expectation as a driving example), consider the following statistical laboratory protocol:\vspace{-0.1cm}
\begin{itemize}
    \item[(i)] A certain microscopic system, called the {\em object system} or {\em object}, is prepared identically (with the same  wave-\newpage function\footnote{ In this chapter, we exclusively deal with pure states. Extending the discussion to mixed states would mainly introduce averages of the considered results across the mixture, with no substantial change in Conclusions 1 to 4 (and 7). We defer this analysis to future research, along with the exploration of open system weak values (derivable from many-body weak values of Appendix B).}) $M$ times in a laboratory.\vspace{-0.1cm}
    \item[(ii)] For each preparation, a standard or so-called ``strong" laboratory measurement of the momentum is registered, resulting in $\x^{(k)}$ for the $k$-th repetition. A standard or ``strong" laboratory measurement is performed by first letting the microscopic object interact with a calibrated ancilliary microscopic system, called an {\em ancilla}, constituting the part of the measurement device that directly interacts with the object. This interaction is designed to correlate in an (approximately) one-to-one fashion the momentum's value of the object with some property of the ancilla, typically its position. In general, the process substantially alters the pre-measurement state of both sub-systems due to the reciprocity of interactions. The position of the ancilla is then coupled in an (approximately) one-to-one fashion with the position of a macroscopic pointer, providing the outcome of the measurement. This last coupling is mediated by the chain of interacting systems that constitute  the measurement device.\footnote{ As a result, in a standard laboratory measurement, the {\em macroscopic pointer} gets correlated in an (approximately) one-to-one manner with the measured property of the {\em object}. In this respect, although the ancilla is always present in experimental terms, one may omit any reference to it, as is generally done. However, the role of the ancilla becomes prominent in what we refer to as ``weak" laboratory measurements in the next subsection. Thus, we already mention the ancilla here in order to clarify the contrast between standard and ``weak" laboratory measurements.} \vspace{-0.1cm}
    \item[(iii)] Finally, the quantity $\langle \x\rangle_M:=\sum^M_{k=1}\x^{(k)}/M$, known as the {\em average value of momentum} is computed.  If $M$ is large enough, upon several repetitions of the protocol with different ensembles of identically prepared systems, we expect the obtained average values to converge to a fixed number $\langle \x\rangle:=lim_{M\rightarrow\infty}\langle \x\rangle_M$. This asymptotic average value is called the {\em (empirical) expectation value of momentum}.\vspace{-0.1cm}
\end{itemize}

 Then, in order to relate this last quantity to the {\em formalism} of quantum mechanics, one should consider the state-vector $|\psi\rangle$ and the momentum operator $\hat P_s$ that models the preparation procedure mentioned above. As is well-known, after assuming the mathematical description of a measurement by \cite{vonNeumann}, including Born's rule, one obtains that $\langle \x\rangle=\langle \psi|\hat{P}_s|\psi\rangle$. Thus, we refer to $\langle \hat{P}_s\rangle:=\langle \psi|\hat{P}_s|\psi\rangle$ as the {\em (formal) expectation value of momentum}. 

Lastly, the {\em ontological} significance of expectation values may vary depending on specific ontological attitudes. Some physicists may argue that expectation values can only be understood as properties of the ensembles, while others may suggest that they also characterize each individual system. Due to the unavoidable backaction of the measurement, some may interpret them as properties of the systems after the measurement has occurred. However, as $\langle \hat{P}_s\rangle$ depends solely on the wavefunction of the systems before the backaction occurs, others may claim that this quantity characterizes the systems'  pre-measurement state. What matters to us is that, regardless of ontological disputes, nobody would question the practical utility of expectations in characterizing quantum systems. As we will see, similar conclusions can be inferred regarding weak values.

\begingroup
\renewcommand{\addcontentsline}[3]{}
\subsection{On the meaning of weak values post-selected in position\vspace{-0.1cm}}

\subsubsection{Weak values as quantities obtained in a laboratory\vspace{-0.2cm}}
\endgroup
\addcontentsline{toc}{subsection}{A. On the meaning of weak values post-selected in position}
\addcontentsline{toc}{subsubsection}{1. Weak values as quantities obtained in a laboratory\vspace{-0.05cm}}

\label{emp1}

Let us now proceed with the experimental description of a weak value, similarly to what we have just done with standard expectation values. We will take the weak value of momentum post-selected in position as a driving example. As we will see, there are two closely related asymptotic-empirical weak values which we will denote by $\w_{e,r}(x,t,\tau)$ and $\w_{e,i}(x,t,\tau)$ respectively.\footnote{The subscript ‘e’ refers to ``empirical", whereas the subscripts ‘r’ and ‘i’ respectively refer to the ``real" and ``imaginary" parts of a complex number. The reason for using these subscripts will be clear after deriving the formal counterparts of the quantities.} Starting with $\w_{e,r}(x,t,\tau)$, this is the quantity that can be approximated after performing the following laboratory protocol and subsequent statistical analysis:
\begin{enumerate}
\item[(i)]  A certain microscopic system, called the {\em object system} or {\em object}, is prepared identically (with the same wavefunction) $M$ times in a laboratory.\vspace{-0.1cm}

\item[(ii)] For each preparation, a ``weak" laboratory measurement of the momentum is registered at the object's time $t$, resulting in a value $\m^{(k)}$ for the $k$-th repetition. In general, a ``weak" laboratory measurement of an observable involves obtaining information about this observable while causing minimal disturbance to the object system. This is usually achieved by coupling the object with the measurement device's {\em ancilla} through a limited interaction that only slightly correlates the targeted object's information (in our case, the momentum) with some property of the ancilla, typically its position. Next, this property of the ancilla is correlated in an (approximately) one-to-one fashion with the position of a macroscopic pointer, as we have described above for a standard laboratory measurement. In this way, the initial (pre-measurement) state of the object is only slightly modified after the interaction.\footnote{Specific instructions for a ``weak" measurement are strongly contingent upon the studied system. The reader can find detailed guidelines for electrons in \cite{displacement} and for photons in \cite{measTrajs}.}

\item[(iii)] A short time $\tau$ after $t$, a standard laboratory measurement of the position of the object is done, yielding the value $x^{(k)}$ for the $k$-th repetition.\vspace{-0.1cm}

\item[(iv)] At this point, $M$ pairs of values $(\x^{(k)}_w, x^{(k)})$ with $k$ in $\{1, ...,M\}$ have been experimentally determined. Next, the average $\x^{(k)}_w$, for those cases in which $x^{(k)}$ is a certain number $x$, is computed. In other words, one computes the average momentum (weakly measured) of those systems found at $x$ a time $\tau$ later. We refer to such a conditional average as the {\em empirical weak value}. Mathematically it can be written as\vspace{-0.05cm}
\begin{equation}
\w^M_{e, r}(x,t,\tau):= \frac{\sum_{k\in\sigma_x} \m^{(k)}}{M_x},
\end{equation}
where $M_x$ is the number of systems found at $x$ and $\sigma_x$ is the set of labels for those systems. The limit of this conditional average when $M\rightarrow\infty$, known as the {\em conditional expectation} in the theory of probability, is what we define as the {\em asymptotic-empirical weak value} $\w_{e,r}(x,t,\tau).$ Note that for a small $\tau$, $\w_{e,r}^M$ is experimentally understood as the {\em local average of momentum around $x$}, being $\w_{e,r}$ its expected value.
\end{enumerate}
Following the theory of probability, for any $\tau$, the conditional expectation $\w_{e,r}$ can be rewritten as $\w_{e,r}(x,t,\tau)=\int d\m\  \m\  \p(\m|x,t,\tau)$, where $\p(\m|x,t,\tau)$ is the conditional probability (density) to obtain the value $\m$, given that the system is found at $x$. That is,
\begin{equation}\label{pcond}
\p(\m|x,t,\tau):=\lim_{M\rightarrow \infty} \frac{\# \text{\small cases: momentum (weakly measured) $\m$ at time $t$, and particles at $x$, a time $\tau$ after}}{M_x}.
\end{equation}
 Finally, in order to relate $\w_{e, r}(x,t,\tau)$ with its counterpart within the quantum formalism, it is useful to express it avoiding probabilities conditioned on $x$. For this, notice that the probability to measure $\m$ as momentum {\em and} $x$ as position, $\p(\m,x\,|\,t,\tau)$, is defined by the equation \eqref{pcond} but replacing $M_x$ for $M$.
Then, given the definition of conditional probability, $\p(\m|x,t,\tau)=\p(\m,x\,|\,t,\tau )/\p(x\,|\,t,\tau)$, with the marginalized probability density $\p(x\,|\,t,\tau)=\int d\m \p(\m,x\,|\,t,\tau)$, one obtains that\vspace{-0.15cm}
\begin{equation}\label{wve}
    \w_{e, r}(x,t,\tau) = \frac{\int d\m\ \;\m\; \mathbb{P}(\m,x\,|\,t, \tau)}{\int d\m\; \mathbb{P}(\m,x\,|\,t,\tau)}.\vspace{-0.15cm}
\end{equation}

Consider now the asymptotic-empirical weak value $\w_{e,i}$. It is defined after a nearly identical protocol, with the exception that, in step (ii) above, instead of the position of the ancilla, it is its momentum (i.e., the conjugate property) that is correlated in an (approximately) one-to-one manner with the macroscopic pointer. As the protocol does not require additional insights, we refer to Appendix A for a more detailed analysis. 

One can, of course, follow the same protocols but perform a ``weak" laboratory measurement of another observable instead of the momentum in step (ii). In the general case, for an observable $G$ we can define the asymptotic-empirical weak values $\g_{e,r}(x,t,\tau)$ and $\g_{e,i}(x,t,\tau)$, after changing $\m$ by the weak measurement of $G$ in the so-far written formulas. At this juncture, we arrive at a clear initial conclusion.
\begin{itemize}
    \item \textbf{CONCLUSION 1:} It is possible to asymptotically determine the weak values $\g_{e,r}(x,t,\tau) $ and $\g_{e,i}(x,t,\tau)$ in a laboratory. For this, averages of ``weak" laboratory measurements concerning $G$ are necessary, using sufficiently large ensembles of identically prepared systems found to be in $x$ a time $\tau$ after the weak measurement.
\end{itemize} 

\subsubsection{Weak values as elements of the quantum formalism\vspace{-0.05cm}}
\label{mat1}

We are now interested in exploring within the formalism of non-relativistic\footnote{Notice that, in more general quantum theories, wherein for instance the Schrödinger equation is no longer applicable, the same empirical weak values might correspond to different formal weak values.} quantum mechanics, what the protocols described in the last sub-section correspond to, regardless of the preferred interpretation or ontology of the theory. Among other basic principles, we will assume the unitary Schrödinger evolution between measurements, the Born's rule and the von Neumann mathematical description of measurements \citep{vonNeumann}. These principles are widely accepted as either fundamental or effective\footnote{For instance, in Bohmian mechanics, the {\em collapse} law is understood to be a consequence of deterministic particle dynamics that “select” an effective wavefunction during the ancilla’s channelization process of a measurement \citep{absolute}.} in practically any non-relativistic quantum theory \citep{vonNeumann, Durr, JordiXavier}.

Let us begin with the formal characterization of the protocol associated with the asymptotic-empirical weak value $\w_{e,r}$. Recall that the procedure involves $M$ repetitions of identically prepared experiments (with $M\rightarrow\infty$). Each repetition comprises, in turn, three steps to be formalized: (a) a weak coupling of the two microscopic systems that we have called the object system, $s$, and the ancilla, $a$; (b) a subsequent standard laboratory measurement of the ancilla’s position; and, (c) a time $\tau$ after, a standard laboratory measurement of the position of the object. Let $x$ be the position coordinate of the object (with corresponding position operator $\hat{X}=\int x|x\rangle\langle x|dx$) and $p_s$ its momentum coordinate (with corresponding momentum operator  $\hat{P}_s=\int dp_s\:p_s|p_s \rangle\langle p_s| $). We will assume with generality that the object’s initial state (at time t) is $\ket{\psi(t)}=\int\psi(p_s)|p_s\rangle dp_s$. Similarly, let $y$ be the position coordinate of the ancilla (with corresponding position operator $\hat{Y}=\int y|y\rangle \langle y|dy$), $p_a$ its momentum coordinate (with corresponding momentum operator $\hat{P}_a=\int p_a|p_a\rangle \langle p_a| dp_a$) and its initial state $\ket{f(t)}=\int f(y)|y\rangle dy$. 

Following \cite{vonNeumann}, we will model standard laboratory measurements in (b) and (c) through the {\em collapse} law and Born's rule, while the ancilla-object interaction in (a) will be modeled through the Hamiltonian $\hat{H}_{sa}=\gamma \hat{P}_s\otimes\hat{P}_a$, which operates during an interval $T$, negligible if compared to the object system's dynamics. This Hamiltonian correlates the object’s momentum with the displacement of the ancilla. As is well-known, the strength of this interaction is given by $\gamma>0$, where larger values of $\gamma$ result in a stronger correlation between the ancilla’s displacement (and eventually the macroscopic pointer’s displacement) and the object’s momentum. Due to the reciprocity of the interaction, the strength $\gamma$ is positively correlated with the disturbance of both systems’ pre-measurement states. Therefore, the weak coupling can be properly modelled assuming that $\gamma$ is small enough. Finally, in order for the macroscopic pointer in (b) to indicate a hypothetical measured value $\x_w$ when the ancilla gets collapsed at $y = \gamma T\x_w$, it is typical to assume that the ancilla’s state is calibrated to be centred around a reference position $y = 0$ (satisfying  $\int y|f(y)|^2dy=0$) with a precision on the order of $\sigma$ (which, formally, implies a support on the order of $\sigma$ for $f(y)$).

Let us now find the formal correlate of $\w_{e,r}$, given the indications made above. Notice that the initial state of the ancilla-object composite system is given by the separable wavefunction $\ket{\Psi(t)}_{sa}=\ket{\psi(t)}\otimes \ket{f(t)}=\int dp_s\: \psi(p_s)|p_s \rangle  \otimes \!\! \int dy \:  f(y) |y\rangle $. Next, during time $T$, the von Neumann interaction among both sub-systems takes place, described by the unitary operator $\hat{U}_{sa}=exp(-i\gamma T\hat{H}_{sa}/\hbar)$. This interaction leads to the following entangled state of the composite system,\vspace{-0.1cm}
\begin{equation}
|\Psi_{pre}\rangle_{sa}:=\hat{U}_{sa}|\Psi(t)\rangle_{sa} = \iint \psi(p_s) f(y-\gamma Tp_s)\:|p_s\rangle\: |y\rangle dp_s dy.
\label{onepm}
\end{equation}
This result holds for any value of $\gamma$, meaning we have not yet derived any consequence of assuming a specific strength for the interaction. At this point, the read-out of the ancilla's position occurs, wherein a measured displacement $y=y_{\m}:=\gamma T \m$ corresponds to an object momentum $\m$. This causes the collapse operation $\hat{I}_s\otimes |y_{\m} \rangle  \langle y_{\m}|$ on the ancilla-object composite system, which leaves this system in the state $\int  \psi(p_s)f({y_{\m}-\gamma T p_s})|p_s\rangle \otimes |y_{\m}\rangle dp_s$, and the object system in the perturbed state
\begin{equation}
|\tilde{\psi}_{\m}(t)\rangle_s := \hat{I}_s\otimes \langle y_{\m}|\Psi_{pre}\rangle_{sa} = \int \psi(p_s) f(y_{\m}-\gamma T p_s)  |p_s\rangle dp_s.
\label{onem}
\end{equation}
This is the object's state after the measurement in (b), for any value of the coupling strength $\gamma$. Thus, the read-out perturbs the object system's wave function differently as a function of the read-out value $y_{\m}$, even if we choose $\gamma$ to be small enough for a weak coupling. The assumption that $\gamma$ is small and thus that we are modelling a ``weak" laboratory measurement finally allows to neglect the terms of second order (and {\em not} of first order, as misleadingly\footnote{ 
For a sufficiently small $\gamma T > 0$, it might seem reasonable to neglect the first-order term as well, assuming that $f(y - \gamma T p_s) \simeq f(y)$. This would imply that the object's state would not be perturbed at all by the weak measurement, $|\tilde{\psi}{\m}\rangle \simeq f(y) |\psi\rangle$ (since $f(y)$ is a factor that disappears through normalization), apparently allowing the possibility to extract information about the object without the individual repetitions experiencing any significant backaction. We argue that such a backaction-free weak measurement is impossible if one wishes to obtain the weak value. The contribution of the zeroth order term to the probability density \eqref{prob} vanishes in the conditional expectation \eqref{intprob}, meaning the weak value emerges strictly from the first order term. Therefore, since the first order term is the key element in the emergence of the weak value, it is also necessary to consider it in the post-weak-measurement state of the system. That is, the correct lowest order approximation for this state is $|\tilde{\psi}{\m}\rangle \simeq f(y) |\psi\rangle+\gamma T df(y)/dy\hat{P}_s|\psi\rangle$, which is a different state as a function of the weakly measured $y$; i.e., the system's perturbation is not negligible.
}
 suggested by some authors \citep{vaidman, weakvalueSingle2}) in the Taylor expansion\footnote{Note that we assumed $1/T\gamma$ to be significantly larger than the length of the support of $p_s$ that the device is measuring.}
\begin{equation}\label{taylor}
 f(y-\gamma T p_s)= f(y)-\gamma T p_s\dv{f(y)}{y}+O((\gamma T)^2).\vspace{-0.2cm}
\end{equation}\newpage
In order to formalize step (iii) in the experimental protocol, we assume that the (slightly perturbed) object system freely evolves for a time $\tau$, according to its time-evolution operator $\hat U_s(\tau) = \exp(-i\hat H_s \tau/\hbar)$, where $\hat H_s$ is the Hamiltonian governing the evolution of the object alone. Finally, after $\tau$, a standard laboratory measurement of the system's position $x$ is performed. According to Born's rule, since we did not normalize the state \eqref{onem}, the joint probability of obtaining the read-out $y=y_{\m}$ in the weak measurement and $x$ in this final standard measurement is $\p(\m,x\,|\,t,\tau)=|\bra{x}\hat U_s(\tau) |\tilde{\psi}_{\m}(t)\rangle_s|^2$. Expanding, we have
\begin{equation}
    \label{prob}
    \p(\m,x\,|\,t,\tau)= \Big|f(y_{\m})\langle x|\hat U_s(\tau)\int \psi(p_s)|p_s\rangle dp_s-\gamma T \dv{f(y_{\m})}{y}\langle x|\hat U_s(\tau)\int \psi(p_s)p_s|p_s\rangle dp_s +O((\gamma T)^2)\Big|^2=\vspace{-0.2cm}
\end{equation}
$$
= |f(y_{\m})|^2|\langle x|\hat U_s(\tau)|\psi(t)\rangle|^2-2\gamma T \Re{f^*(y_{\m})\dv{f(y_{\m})}{y} \langle \psi(t)|\hat U^\dagger_s(\tau)|x\rangle\langle x|\hat U_s(\tau)\hat{P}_s|\psi(t)\rangle} +O((\gamma T)^2).
$$
If the ancilla's rest state $f(y)=R(y)exp(iS(y)/\hbar)$ has an odd phase $S(y)$ and an odd or even modulus $R(y)$, implying that $\int yf^*(y)\dv{f(y)}{y} dy=-1/2$ and $\int y |f(y)|^2dy=0$, we obtain\vspace{-0.2cm}
\begin{equation}
    \label{intprob}
    \int \m\:\p(\m,x\, |\, t,\tau)d\m = \frac{1}{\gamma T}|\langle x|\hat U_s (\tau) |\psi(t)\rangle |^2\Re{\frac{\langle x|\hat U_s(\tau)\hat{P}_s |\psi(t)\rangle}{\langle x|\hat{U}_s(\tau)|\psi(t)\rangle}}+O(1).
\end{equation}
At this point, one should realize the significance of the first-order term in \eqref{taylor}, as it constitutes the dominant term. 

Lastly, for the marginalized probability density $\int\p (\m,x\,|\,t,\tau)d\m$, we obtain
\begin{equation}\label{marginalized}
    \int\p (\m,x\,|\,t,\tau)d\m = \frac{1}{\gamma T}|\langle x|\hat U_s(\tau)|\psi(t)\rangle|^2 +O(1),
\end{equation}
from which we deduce that, to leading order, the asymptotic-empirical weak value $\w_{e,r}$ in \eqref{wve} within the formalism of quantum mechanics yields\vspace{-0.3cm}
\begin{equation}\label{wvmre}
    \w_{e,r}(x,t,\tau)  \simeq \Re\Bigg[\frac{\langle x|\hat U_s(\tau)\hat{P}_s |\psi(t)\rangle}{\langle x|\hat{U}_s(\tau)|\psi(t)\rangle}\Bigg].
\end{equation}
 As we show in Appendix A, a similar derivation (where the same assumptions about the interaction strength $\gamma$ and the ancilla's wavepacket $f(y)$ are adopted) now for the laboratory protocol defining $\w_{e,i}$, leads to
\begin{equation}\label{wvmim}
    \w_{e,i}(x,t,\tau)\simeq \Im\Bigg[\frac{\langle x|\hat U_s(\tau)\hat{P}_s |\psi(t)\rangle}{\langle x|\hat{U}_s(\tau)|\psi(t)\rangle}\Bigg].
\end{equation}
This strongly suggests the definition of a {\bf \em formal weak value} encoding the two asymptotic-empirical weak values,\vspace{-0.1cm}
\begin{equation}
\w_{f}(x,t,\tau):=\frac{\langle x|\hat U_s(\tau)\hat{P}_s |\psi(t)\rangle}{\langle x|\hat{U}_s(\tau)|\psi(t)\rangle}\simeq\w_{e,r}(x,t,\tau)+i\:\w_{e,i}(x,t,\tau).
\end{equation}
In general, if in the laboratory protocol we had weakly measured another observable $G$ with an associated operator $\hat{G}_s$, and defined the asymptotic-empirical functions $\g_{e,r}(x,t,\tau), \g_{e,i}(x,t,\tau)$ analogously, the resulting formal weak value would be given by\vspace{-0.4cm}
\begin{equation}\label{wvf}
    \g_{f}(x,t,\tau):=\frac{\langle x|\hat U_s(\tau)\hat{G}_s |\psi(t)\rangle}{\langle x|\hat{U}_s(\tau)|\psi(t)\rangle}\simeq\g_{e,r}(x,t,\tau)+i\: \g_{e,i}(x,t,\tau).\vspace{-0.1cm}
\end{equation}
Note that formal weak values do not depend on any particular measurement context and contain information exclusively about the state of the system before its measurement, $\ket{\psi(t)}$, similarly to what occurs with expectation values. Thus, we reach the following conclusion:
\begin{itemize}
    \item \textbf{CONCLUSION 2:} The experimental conditional expectations of Conclusion 1, for a weak enough measurement, can be predicted {\em within any non-relativistic }quantum theory (where the von Neumann measurement protocol and the Born rule are accepted), through the formal weak value \eqref{wvf}, which only depends on the  state $|\psi(t)\rangle$ of the pre-measurement system and its propagator $\hat{U}_s(\tau)$. 
\end{itemize}
Weak values for which $\tau$ is very small can be understood as {\em expectation values around the position $x$}, both formally and experimentally. We have already seen that this is the case experimentally.  
To show this is true also formally, let us define the limit weak value $\g_\ell(x,t):=\lim_{\tau\rightarrow 0} \g_f(x,t,\tau)$, which is a simultaneously well-defined field in configuration and time for any observable $G$ (even for those not commuting with $\hat X$). This field satisfies\vspace{-0.1cm}
\begin{equation}\label{dens}
    \int |\psi(x,t)|^2\: \g_\ell(x,t)\: dx=\langle \psi(t) |\hat{G}_s|\psi(t)\rangle=\langle \hat{G}_s\rangle(t),\vspace{-0.1cm}
\end{equation}
where we have defined $\psi(x,t):=\langle x|\psi (t) \rangle$. That is, the standard expected value of an operator $\hat{G}_s$ can be recovered through the spatial average of its limit weak value $\g_\ell(x,t)$. This is why it is appropriate to refer to  $\g_\ell(x,t)$ as the {\bf \em local expectation value of $G$ (around x)}. It provides {\em information} on how the expectation value $\langle \hat{G}_s\rangle(t)$ is distributed along the configurations $x$. Notice that, we use the ambiguous term \textit{``information"} here deliberately, because we do not want to attach any ontological significance to the local expectations, beyond their experimental and formal interpretations. Moreover, since $\hat{G}_s$ is Hermitian, even the real part of the local expectation field alone, $\g^r_{\ell}(x,t):=\Re [\g_\ell(x,t)]$, behaves itself as an expectation density: $    \int |\psi(x,t)|^2\: \g^r_{\ell}(x,t)\: dx=\langle \hat{G}_s\rangle(t).$ This implies that the complex part of the weak value, with generality, satisfies $   \int |\psi(x,t)|^2 \Im [\g_\ell(x,t)]dx = 0$.\footnote{The attentive reader may have noticed that, when referring to $\g_\ell(x,t)$ as local expectations, we are borrowing the terminology introduced by \cite{Holland}. We are willingly doing so but two differences should be noted. First,  only $\g^r_\ell$ is defined to be a ``local expectation" in \cite{Holland}. Second, Holland introduces local expectation values within an ontological discourse, interpreting them as properties of individual Bohmian particles and without a clear reference to any laboratory protocol.} 
Summing up, we can conclude:

\begin{itemize}
    \item \textbf{CONCLUSION 3:} From both an empirical perspective and a formal perspective {\em within any non-relativistic quantum theory}, weak values define expectation values around certain configurations $x$ when $\tau\rightarrow 0$, known as {\em local expectation values}. Either the complex field $\g_\ell(x,t):=\lim_{\tau\rightarrow 0} \g_f(x,t,\tau)$ or the real field $\g^r_{\ell}(x,t):=\lim_{\tau\rightarrow 0} \Re [\g_f(x,t,\tau)]$ are possible local expectation values.\vspace{-0.2cm}
\end{itemize}

Assuming the polar decomposition of the object system’s wave function, $\psi(x,t)=R(x,t)e^{iS(x,t)/\hbar}$, local expectation values can be further analysed. For instance, the local expectation value of the momentum can be developed as\vspace{-0.3cm}
\begin{equation}
\w_\ell(x,t) =  \frac{ \langle x| \hat P_s|\psi(t)\rangle }{\langle x|\psi(t)\rangle}=\frac{\partial S(x,t)}{\partial x}-i \frac{\hbar}{R(x,t)}\frac{\partial R(x,t)}{\partial x}.
\label{wvm}
\end{equation}
It turns out that this example vividly highlights the informational richness of weak values. If the real part of this local expectation is known, which is experimentally given by  $\lim_{\tau\rightarrow 0}\w_{e,r}(x,t,\tau)$, it is possible to compute the phase of the wavefunction $S(x,t)$ by its integration, up to a constant (a global phase). Moreover, in the laboratory protocol to obtain it, we also determine the probability density $R^2(x,t)$ through the marginal density $\p(x\,|\,t,\tau)$ of equation \eqref{marginalized}. This implies that the laboratory protocol for $\w_{e,r}$ (with a small enough $\tau$), reveals the wavefunction of the systems as it was prior to any measurement back-action, $|\psi(t)\rangle$.\footnote{Again, we are assuming here a single-particle quantum system in a pure state.}$^,$\footnote{In fact, this is merely one among numerous ways in which weak values enable the experimental determination of the wavefunction of a system. For example, in the pioneering experimental work by \cite{directWF}, the authors determined the wavefunction through {\em momentum post-selected} local expectations (a class of weak value we will elucidate in Section C).} It follows that weak values and their functions can be as informative regarding the characterization of quantum systems as the wavefunction itself is. Thus, we can conclude:
\begin{itemize}
    {\item \textbf{CONCLUSION 4:} The weak values of $G$ and their associated protocols provide new ways to characterize quantum systems, beyond the corresponding operator's expectation values.}
\end{itemize}

Conclusions 1 to 4 suffice to convey one of the main messages of this paper, particularly aimed at physicists leaning towards an empiricist attitude: weak values are useful tools that are worth considering and further developing, regardless of discussions about their ontological significance. Let us make, next, a brief incursion into interpretive matters, addressing the ontological significance of weak values according to different quantum theories.

\subsubsection{Weak values as elements of reality}
\label{onto}\vspace{-0.15cm}

 Presumably, each quantum theory provides a distinctive view on the ontology of weak values. Let us first consider what we refer to as the “standard” or “orthodox” view, that attributes properties to microphysical systems via the, so-called, “\textit{eigenvalue-eigenstate link"}. As it is well-known, this link establishes that a system has a property \textit{G} with determinate value $g$ if and only if the system’s wavefunction is an eigenstate of the operator representing the property, $\hat{G}_s$, with eigenvalue $g$. As a consequence of this interpretive tenet, if the wavefunction of the system is a superposition of eigenstates of $\hat{G}_s$ associated with different eigenvalues, simply, the system does not determinately possess the property \textit{G}. This view also includes the collapse-upon-measurement as one of its postulates. Given that the collapse ensures that, after a measurement of property \textit{G}, the wavefunction transforms into an eigenstate of the operator $\hat{G}_s$, it follows from the standard view that, \textit{after }a measurement of property \textit{G}, this property is always a determinate, well-defined property of the system with the value corresponding to the measured outcome. Yet this is \textit{not} the case, in general, \textit{before} the measurement. So the standard view has the consequence that, in general, measurement processes not only disturb the measured system but “create” or make determinate the measured property \citep{Mermin}.\vspace{-0.15cm}

Regarding weak values, when the state of the system $\ket{\psi}$ is already an eigenstate of the observable $\hat{G}_s$, with $\hat{G}_s|\psi\rangle=g|\psi\rangle$, its associated formal weak value gives $\g_f(x,t,\tau)=g$ for all $x$ and $\tau$. In light of the above explanation, within the standard view, such a system determinately has the property G with value $g$, regardless of its position or anything else. Therefore, even if $\g_f(x,t,\tau)$ is computed through an average, the standard view attributes the weak value, in such a case, to each individual system of the ensemble (as the well-defined property \textit{G} with value $g$). \vspace{-0.1cm}

However, in general, the state $\ket{\psi}$ is \textit{not} an eigenstate of $\hat G_s$. In such case, according to the eigenvalue-eigenstate link, the determinate properties of the system would correspond to a set operators\textit{ different than} $\hat G_s$. Given the scarcity of properties available to the advocate of the standard approach, we can conclude: 
\begin{itemize}
    \item \textbf{CONCLUSION 5:} Quantum theories assuming the eigenstate-eigenvalue link as a fundamental interpretive postulate cannot generally relate weak values with properties determinately possessed by the system (nor with their averages) and thus these theories have a too sparse ontology to attribute an ontological meaning to weak values.\vspace{-0.1cm}
\end{itemize}
One may wonder what would happen if we understood that the average observable $G$ (such as momentum or energy) around a position $x$ was actually a property of any of the systems found at $x$. This is what happens according to the standard view, \textit{but only for systems whose state  is an eigenstate} of $\hat G_s$. Now we want to drop this restriction. In such an alternative view, the local expectation value of the momentum would imply a well-defined momentum for the particle system at all times (determined by the system's position). Consequently, the particle of the system would trace well-defined trajectories in $x$ space. A minimal version of this theory would focus on the real local expectations $\g^r_\ell(x,t)$, which already yield the operator expectations. According to such a minimal version, the momentum of the particle when it is at $x$ would be equal to the gradient of the phase of the wavefunction, as shown in equation \eqref{wvm}. For well-behaved wavefunctions, this would result in non-crossing deterministic trajectories, each ascribable to individual realizations of the system \citep{nonCrossing}. This, in turn, would imply that if we measured at time $t$ the position of the system with trajectory $x^{k}(t)$, the result would correspond with $x=x^{k}(t)$, because that would have been the actual position of the particle. However, after the measurement, the trajectory would be governed by the phase of a wavefunction that would have been uncontrollably perturbed,\footnote{We saw previously that the perturbation is significant even for the weakest of the measurement couplings.} so the entire individual trajectories, although determinate at the ontological level, would not be epistemically accessible. Hence, a stochastic description of positions would still be necessary epistemologically, to have an empirically adequate theory. Similarly, the other observables would also be ontologically determined through the position of the particles. In this theory, the real local expectation of $G,$ $\g^r_\ell(x,t)$, would give the value of $G$ corresponding to the particle when found to be in $x$ at time $t$. 
It is worth noting that, in this theory, an eigenstate of an operator is a state in which all possible trajectories exhibit the same value for the corresponding property. This would explain why, when we measure an observable in an eigenstate, we always obtain the same value with a probability equal to 1.

Such a ``literal interpretation of local expectations" is indeed very close to Bohmian mechanics \citep{bohm, Durr, Holland, JordiXavier}, with the primary difference being the terminology used for certain functions. For instance, in both cases, the momentum of the particle, denoted as $p_B(x, t)$, is equal to (the real part of) the momentum weak value. However, the kinetic energy is given by $p_B^2(x, t)/(2m)$ in Bohmian mechanics, while, in the ``literal interpretation of local expectations", it is the real part of the weak value of $\hat{P}_s^2/(2m)$ that is referred to as the kinetic energy. As we will see in the next subsection, these quantities do \textit{not} coincide. In the Bohmian theory, the difference between them is the so-called  “quantum potential”.

Given that, in Bohmian mechanics,  $p_B = \frac{\partial S(x,t)}{\partial x}$ is the pre-existent value (previous to any interaction) of the momentum of the particles, and we have demonstrated that this quantity is equivalent to the real local expectation of momentum, $p_B = \Re(\w_\ell)$, it follows that the protocol to obtain the weak value $\Re(\w_\ell)$ described in the subsection \ref{emp1} serves as an experimental protocol for accessing this property within the framework of Bohmian mechanics. Now, it is essential to emphasize that, according to alternative quantum theories, the same experimental protocol may be interpreted differently because, among other things, it involves averaging and post-selecting data obtained from an ensemble of different experiments, rather than from an individual experiment. Furthermore, there exist non-standard but empirically equivalent versions of Bohmian mechanics that propose different velocity fields \citep{Deotto}. In these theories, it is evident that the protocol associated with $\Re(\w_\ell)$ does \textit{not} constitute a measurement in any sense of the momentum of the particles. This shows that, to the extent that one can talk of ``measurement" in this context, it is a heavily theory-laden notion. Moreover, since all these theories are empirically equivalent, one can, at most, empirically establish what they have in common but not where they diverge.\footnote{Thus, the actual existence of the Bohmian trajectories cannot be experimentally proved. For a more in-depth discussion on the ``undetectability" of Bohmian trajectories, please refer to \cite{frank} in this volume.} It is the ontology of each quantum theory that determines the ontic import of each weak value.

Finally, let us remark that there is a theory where local expectations correspond to {\em averages} of ontologically determined properties that may vary from preparation to preparation. According to this theory, the particles follow well defined trajectories that evolve stochastically and the Bohmian trajectories represent average paths. This theory is Nelson’s stochastics mechanics where the expected (but not the individual) particle momentum is equal to the Bohmian momentum $p_B$ (accompanied by an {\em osmotic momentum} component describing a diffusive process, given by $p_O:=\Im{\w_\ell}$). Of course, as it happens in Bohmian mechanics, there might be a difference among Nelson’s theory and a “literal interpretation of weak values” that naively takes the weak value of any operator, $\hat{G}$, to represent the average of the property $G$ in the ensemble. 

We can therefore conclude that:\vspace{-0.05cm}
\begin{itemize}
    \item {\bf CONCLUSION 6:} Bohmian mechanics {\em allows} to interpret weak values post-selected in position (when $\tau\rightarrow 0$) as properties of individual systems at specific positions.\footnote{ We have pointed out that within Bohmian mechanics, in general, one can think of the real local expectation $\g^r_\ell(x,t)$ associated with the operator  $\hat G_s$ as representing some property (not necessarily $G$) of the systems located at position $x$ at time $t$. This interpretation is reinforced by prominent Bohmians, such as \cite{undivided} or \cite{Holland}. Other prominent Bohmians, like \cite{bell} or \cite{naiveOperators}, consider that the only property attributed to the Bohmian particles is their position, because it suffices to explain all experiments in the laboratory (as they ultimately involve only the positions of measurement apparatus pointers). Moreover, advocating for a minimal set of properties eliminates the possibility of drawing misleading conclusions. Nonetheless, a broader number of properties provides practical heuristic power in characterizing quantum systems, as we show in the case study. } In Nelson's stochastic trajectory theory, these weak values {\em can} instead be understood as averages of individual system properties.
\end{itemize}

Note that other quantum theories may result in different, alternative ontological readings of weak values. One may just claim, for instance, that the momentum’s local expectation value, \eqref{wvmre}, is merely ``the gradient of the phase of the wavefunction", and that this is what is measured when performing the corresponding laboratory protocol \citep{fankhauser}. We do not intend to object to such analyses, but in our view, they remain incomplete in terms of characterizing the ontological meaning of the local expectation value unless they are accompanied by a characterization of the \textit{ontological} meaning of the wave function itself. This can potentially give rise to further challenges, particularly considering the multi-dimensional nature of the wave function field.\footnote{Notice that Bohmian-like theories avoid this difficulty, at least, when it comes to specifying the nature of weak values. For these theories, weak values unambiguously represent properties of the \textit{particles}, even though they formally depend on the wave function.} However, due to space constraints, we have limited our discussion of the ontological meaning of weak values to the alternatives already presented.\vspace{-0.2cm}
\begingroup
\renewcommand{\addcontentsline}[3]{}
\subsection{On the meaning of a function of a weak value\vspace{-0.1cm}}
\endgroup
\addcontentsline{toc}{subsection}{B. On the meaning of a function of a weak value}

Weak values are interesting, among other reasons, because usually they cannot be rewritten as the expectation of a Hermitian operator, thus offering an additional characterization tool for quantum systems. It turns out that a {\em function} of weak values (such as $p_B^2$,  that is, the square of the Bohmian momentum $p_B=\Re{\w_\ell}$ in equation \eqref{wvm}), in addition to not being expressible as a standard expectation either, generally cannot be written as a weak value of some Hermitian operator. Not at least of the most obvious choice of operator (for example $\hat{P}_s^2$). And yet, experimentally, \textit{functions} of weak values can still be obtained in a laboratory simply by applying those functions to the numbers resulting from the previously explained weak value protocols. This means that {\em functions} of weak values constitute a vast source of new information to characterize quantum systems, even beyond weak values themselves. As we will show in the case study, far from being arbitrary operations performed on the experimental results, they can posses very interesting features.

\newpage
Let us illustrate this with a particular example (that relates with the discussion of section II.A.3 and the case study). In terms of the polar decomposition of the wavefunction, the local expectation of the kinetic energy operator $\hat{P}_s^2/2m$ (the {\em weak value of the square of the momentum operator}) reads:\vspace{-0.05cm}
\begin{equation}
     \frac{1}{2m}\frac{ \langle x| \hat P_s^2 |\psi(t)\rangle }{\langle x|\psi(t)\rangle} = \frac{1}{2m}\qty(\pdv{S(x,t)}{x})^2-\frac{\hbar^2}{2m}\frac{1}{R(x,t)}\pdv[2]{R(x,t)}{x}-i\frac{1}{2m}\qty[\hbar \pdv[2]{S(x,t)}{x}+2\frac{\hbar}{R(x,t)}\pdv{R(x,t)}{x}\pdv{S(x,t)}{x}].\label{p2}
\end{equation}

If one re-examines the momentum local expectation $\w_\ell=p_B+i\,p_O$ developed in \eqref{wvm} (which encodes the Bohmian momentum $p_B={\partial S(x,t)}/{ \partial x}$ and the Nelsonian osmotic momentum $p_O=-(\hbar/R){\partial R(x,t)}/{ \partial x}$), it becomes evident that the {\em square of the momentum operator's weak value}, $\w_\ell^2/2m$, yields a different thing. 
The real part of \eqref{p2}, which is enough to recover the kinetic energy's {\em standard expectation}, contains indeed the {\em square of the momentum operator's real weak value}, $\Re{\w_\ell}^2/2m$, namely, the Bohmian kinetic energy, but there is an additional term that depends on the osmotic momentum $p_O$. This term is called the \textit{quantum potential} in Bohmian mechanics,\vspace{-0.05cm}
\begin{equation}
    Q(x,t):=-\frac{\hbar^2}{2m}\frac{1}{R(x,t)}\pdv[2]{R(x,t)}{x}=\frac{1}{2m}\qty(\hbar\pdv{p_O(x,t)}{x}-p_O(x,t)^2).\vspace{-0.05cm}
    \label{qpot}
\end{equation}
Such a development shows that it is possible to determine $p_B^2/2m$ and $Q$ in a laboratory as {\em functions} of the local expectation of momentum $\w_\ell$. Our point is, as we will show in the case study, that the individual quantities $p_B^2/2m$ and $ Q $ separately provide a richer source of information than their sum, the kinetic energy's real local expectation.
\begin{itemize}
    \item {\bf CONCLUSION 7:} In general, {\em functions of weak values of operators} are not equivalent to {\em weak values of functions of operators}. However, {\em functions} of weak values are still experimentally determinable by applying those functions to empirically determined weak values (e.g., when post-processing). Therefore, they constitute predictive tools that provide an additional contrastable source of information about quantum systems.
\end{itemize}
It is important to note that we have not addressed any ontological implication of the {\em functions} of weak values. Therefore, conclusion 7 remains significant for an empiricist attitude reader.\vspace{-0.15cm}

\subsection{Generalization to other post-selections\vspace{-0.1cm}}
If we keep the same state preparation procedure for the empirical weak value protocol and perform a weak measurement of the observable $\hat{G}_s$ as previously described, but instead of measuring the position of the system particles $x$ after time $\tau$, we measure another observable $B$ with associated operator $\hat{B}=\sum_b|b\rangle \langle b|$, we can define additional weak values. These are known as weak values ``post-selected in $B$". Empirically, they correspond to the average of weak measurements for the replicas in which a specific eigenvalue $b$ is obtained in the subsequent measurement. The derivation of the corresponding formal weak values is analogous to the one that we have discussed, and it leads to the following expression for the generalized formal weak values\vspace{-0.1cm}
\begin{equation}
    \g_f^B(b,t,\tau)=\frac{\langle b|\hat U_s(\tau)\hat G_s|\psi(t)\rangle}{\langle b|\hat U_s(\tau)|\psi(t)\rangle}.\vspace{-0.1cm}
\end{equation} 
When the limit $\tau\rightarrow 0$ is considered, both empirically and formally, the standard expectation $\langle \hat{G}_s\rangle(t)$ can still be recovered by an average of the weak values over all possible $b$, weighted by their probability density $|\psi(b,t)|^2:=|\langle b|\psi\rangle|^2$. Thus, both empirically and formally, all the Conclusions 1 to 4 (even 7) can be generalised, since $\g_\ell^B(b,t)=\lim_{\tau\rightarrow 0}\g_f^B(b,t,\tau)$ is still a local expectation value, but now in the ``locality" of $B=b$, whatever this might mean.

The standard approach with the eigenvalue-eigenstate link has the same difficulties to generally attribute an ontological meaning to weak values post-selected in \textit{B }(or their corresponding ``local" expectation values) as it has in the case of weak values post-selected in position. In fact, even position-based ``hidden" variable theories, such as Bohmian mechanics, do not attach a natural ontological meaning to them. Following the same ``literal" reading of weak values however, one can define a so-called ``modal" theory for each post-selection $B$, that would do so. These theories would have a ``hidden" variable ontology in the usual sense but with well-defined material trajectories in the \textit{b}-space instead of $x$-space. According to them, quantities such as  $\g_\ell^B(b,t)$ would constitute determinate properties of the individual systems ``located" at $B=b$, and the corresponding laboratory procedures would be the natural protocols to determine those properties. 

While general weak values may be valuable in characterizing quantum systems, we consider that position post-selected weak values hold a privileged status. Position is epistemically privileged since one can consider that all measurements are ultimately measurements of position (the position of a pointer, pixel on a screen, or similar indicators). But position is special in several other respects. For instance, in the position basis, the Schrödinger equation has a natural decoupling into a continuity equation and the well-known Hamilton-Jacobi equation (or even into a version of Newton’s second law) \citep{bohm, Holland}.  The resemblance of these equations with the classical ones, allows to extend the intuitions and methods of classical fluid and particle dynamics to the quantum domain in a \textit{guided} and rigorous manner.

\section{Exemplifying Bohmian-like theories as Heuristic Tools}
\label{casestudy}
Finally, let us present a case study to illustrate our principal theses: the utility of weak values and their {\em functions} as novel ways to characterize quantum systems, the fact that this utility is independent of assessments of their ontological import, and how their natural incorporation into Bohmian-like theories make these theories practically valuable tools. Our case study will delve into a cornerstone issue in quantum thermodynamics: the thermalization of a closed quantum system. To maintain clarity, we will provide a qualitative overview of the problem and highlight the relevant results for our argument. A more detailed technical analysis of this case study can be found elsewhere \citep{thermalizationCX}.

\begin{figure}[h!]
\includegraphics[width=0.81\linewidth]{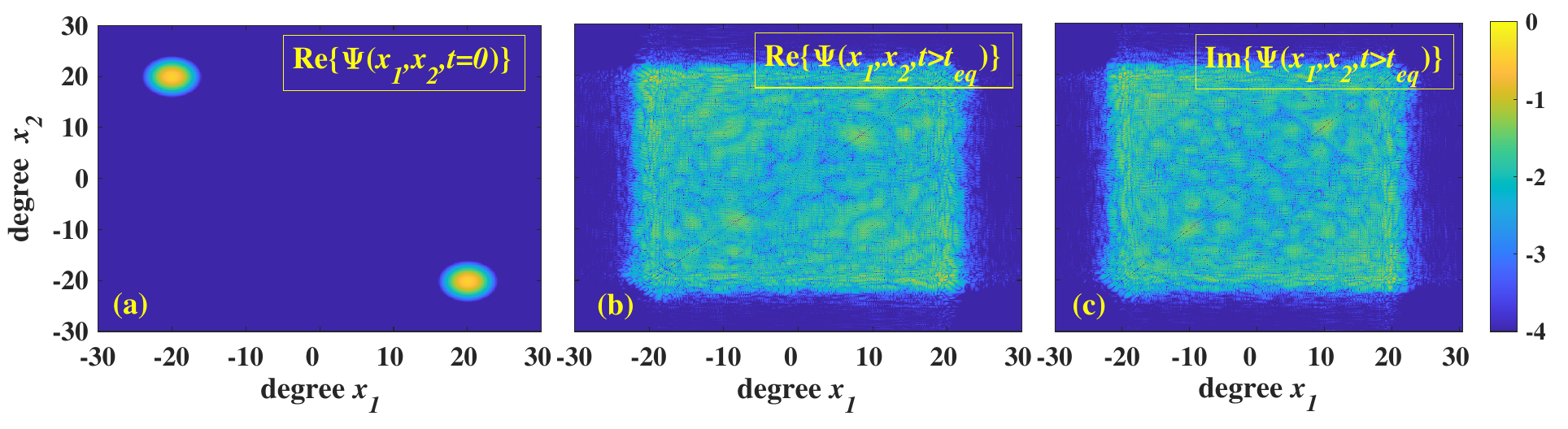}
\caption{Time evolution of a two-particle wavefunction $\Psi(x_1,x_2, t)$ in a disordered harmonic potential. (a) Real part of the wavefunction at $t=0$. (b) Real part of the wavefunction at a representative time $t>t_{eq}$. (c) Imaginary part at the same time $t>t_{eq}$ (notice that the imaginary part at t=0 is zero for this particular scenario). All vertical axes and colormaps represent the same ranges. Atomic units are used. The colormap is the $log_{10}(\cdot)$ value of the modulus of each component.\vspace{-0.1cm}}
\label{fig2}
\end{figure}

Let us consider two\footnote{In order to provide a more concise discussion of weak values, we have so far introduced them for quantum systems with only one
degree of freedom. In Appendix B we generalize weak values to many-body quantum systems with an arbitrary number of degrees of
freedom.} identical electrons (with respective degrees of freedom $x_1$ and $x_2$) confined in a disordered harmonic trap with Coulomb and exchange interactions between them. In general, after multiple scattering events with the speckles of a random pattern in the (disordered) potential field, an initially localized wavepacket unitarily evolves towards a {\em fully dispersed} state. This time evolution is illustrated in Figure \ref{fig2}. It is natural to infer that an \textit{equilibrium} state is eventually reached. Indeed, if we calculate the standard expectations of observables such as the kinetic energy or the velocity, we will observe that they relax over time, converging to steady-state values. The characteristic time constant that quantitatively describes this equilibration is called {\em the thermalization time}, $t_{eq}$. \vspace{0.1cm}

\begin{figure}[h!]
\includegraphics[width=0.79\linewidth]{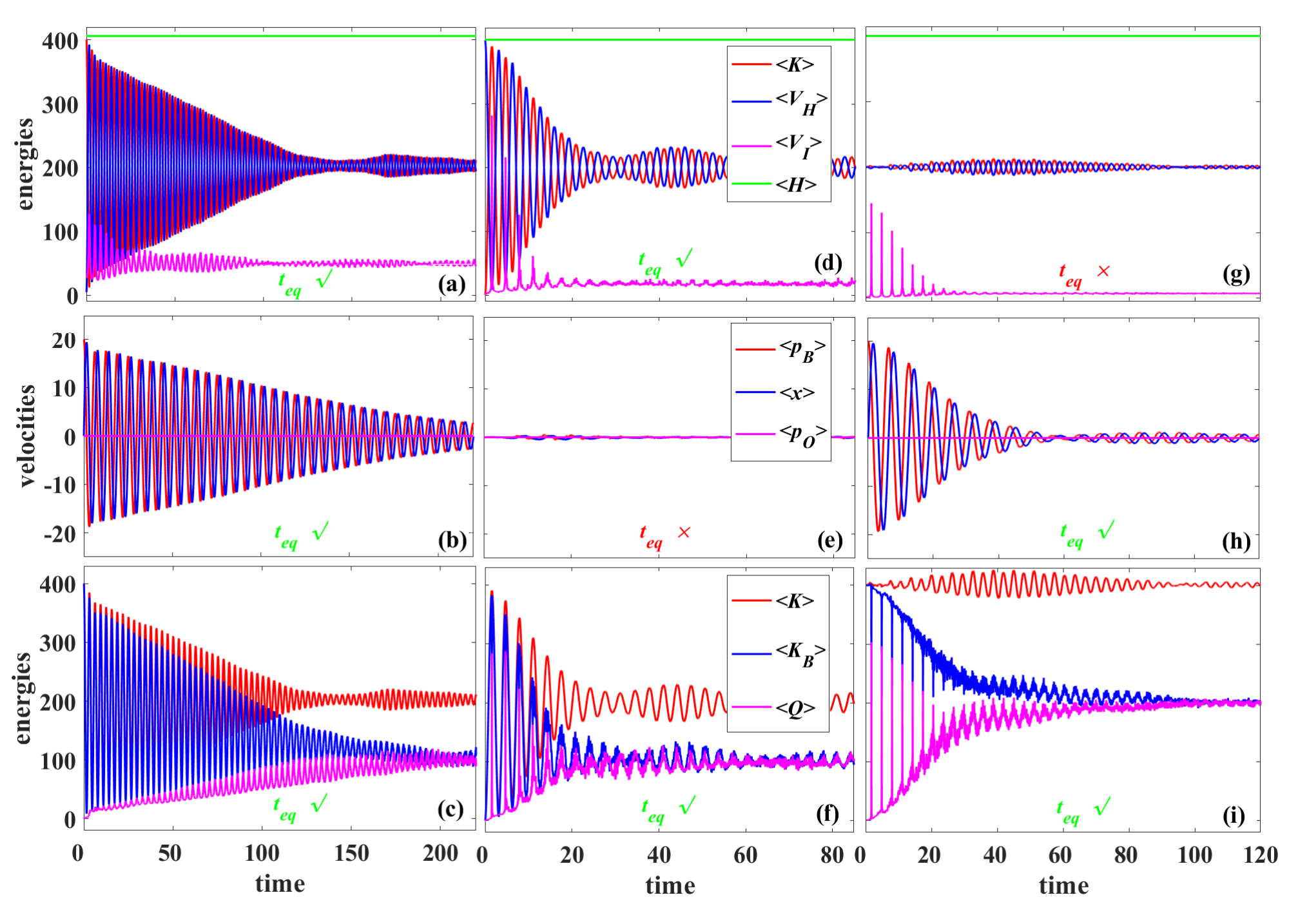}
\caption{In the first column we show a representative initial condition where energy, position and momentum expectations serve to identify thermalization. In the second column, a scenario in which position and momentum are ``blind" to the onset of thermalization. In the third column, a scenario where kinetic and potential energies are unable to reflect thermalization times. The first row shows the time evolution of standard energy expectations: the standard kinetic $\langle K \rangle$, harmonic $\langle V_{H} \rangle$, Coulomb repulsion $\langle V_{I} \rangle$, and total $\langle H \rangle$ energies ($\langle V_{I} \rangle$ is magnified by $100$). The second row shows the time evolution of standard momentum and position expectations $\langle p_B \rangle, \langle x \rangle$, together with the osmotic momentum expectation $\langle p_O \rangle$. The third row shows the time evolution of Bohmian $\langle  K_B \rangle$ and osmotic $\langle Q \rangle$ kinetic energy expectations, relative to $\langle K \rangle$. Values in (g) are divided by two. Legends in left and right columns follow the ones in the middle column. Vertical axis in middle and right columns are the same as in left column. All in atomic units.\vspace{-0.05cm}}
\label{fig1}
\end{figure}

In Figure \ref{fig1}, three different two-particle simulations are presented to exemplify various scenarios in such a quantitative study of thermalization. The left column depicts a case where both particles have initial wave packets with central positions close to zero and identical central momenta. In the middle column the initial wave packets have zero central momentum but opposite central positions (this is the situation depicted in Figure \ref{fig2}). In the right column, the initial central positions are opposite and the central momenta are initially large. 

The process of thermalization is revealed in Figure \ref{fig1} through the time evolution of various expectation values. In the top row, Figures \ref{fig1}(a) and (d) depict the gradual evolution of kinetic and potential energy expectations towards thermalization. In the middle row, Figures \ref{fig1}(b) and (h) display the thermalization of position and velocity expectations.\vspace{-0.1cm}

Our argument revolves around the observation that in certain configurations, many standard expectation values fail to capture the thermalization time due to conservation symmetries. This is illustrated in Figure \ref{fig1}(e), where position and velocity expectations become ``blind" to thermalization, or in Figure \ref{fig1}(g), where standard energy expectations are the ``blind" ones. Fortunately, there exist decompositions of standard expectations into expectations of {\em weak value functions} that still convey the thermalization signature for most such scenarios. This already highlights the practicality of weak values beyond conventional quantum metrics and any interpretative or ontological discussion. Yet, not all decompositions into weak value functions  successfully characterize the thermalization time. And here is our main point: far from needing to look for informative weak values by trial-and-error, Bohmian-like theories naturally suggest appropriate decompositions due to their close connection to some classical notions.\vspace{0.05cm}

Let us illustrate this through particular decompositions of the {\em kinetic energy} that naturally emerge from Bohmian-like theories. In particular, we present the analyses suggested by Bohmian mechanics, the hydrodynamic quantum theory \citep{Madelung} and Nelson's stochastic mechanics.

Within Bohmian mechanics, along the direction $j$, all the energy contributions apart from the classical potential energy, are attributed to the Bohmian kinetic energy, $K_{B,j}=p_{B,j}^2/2m$, and the quantum potential $Q_j$ (as defined in \eqref{qpot} for a single-particle system). In terms of standard quantum operators, the energy contribution along $j$ that is not due to the classical potential, is given by the kinetic energy operator $\hat{P}_j^2/2m$. As we found in equation \eqref{p2}, the (real) local expectation of this operator is equal to $K_{B,j}+Q_j$, which implies the decomposition $\langle \hat{P}^2_j/(2m)\rangle=\langle K_{B,j}\rangle+\langle Q_{j}\rangle$. We also found that the two quantities in the right hand side, although natural in Bohmian mechanics, are not individually associated with Hermitian operator expectations. They are {\em functions of weak values}, for which we still devised a laboratory protocol. Namely, by following the empirical protocol for local expectations, one can determine both the local expectation $p_{B,j}$ and the probability density $R^2(x,t)$, to subsequently evaluate $K_{B,j}$ and $Q_j$ employing their respective functions.  

This decomposition remains natural in hydrodynamic versions of quantum mechanics, where possible Bohmian trajectories can be viewed as elements of a fluid. In these formulations, the quantum potential $Q_j$ arises from the internal pressure of the fluid, and thus, it is treated in analogy to classical fluid mechanics as part of the total kinetic energy of the fluid elements, along with their momentum-associated energy $K_{B,j}$.\vspace{-0.05cm}

A similar, yet alternative decomposition is also natural in Nelson's theory, where the real and imaginary parts of the momentum local expectation are the two momentum components of the stochastic trajectories, respectively, the current momentum $p_{B,j}(\vec{x},t)$ and the osmotic (or diffusion) momentum $p_{O,j}(\vec{x},t)$. Two momentum contributions imply two kinetic energy contributions, one being $K_{B,j}$ and the other one being $K_{O,j}:=p_{O,j}^2/2m$. It is shown in \cite{thermalizationCX} that these functions are related to the standard kinetic energy expectation by $\langle \hat{P}^2_j/(2m)\rangle=\langle K_{B,j}\rangle+\langle K_{O,j}\rangle$.\footnote{This means that $\langle K_{O,j}\rangle=\langle Q_{j}\rangle,$ although $K_{O,j}\neq Q_j$, as discussed in \cite{thermalizationCX}.} Hence, we can alternatively decompose the (standard) average kinetic energy into these two functions of weak values, which are likewise experimentally determinable.\vspace{-0.1cm}

Now, these ``hidden"-variable theories not only serve as a means to derive such decompositions, but within their conceptual frameworks, it is natural that they are valid decompositions for identifying the thermalization time. It is inherent to any of the three approaches that, following the classical concept of {\em energy equipartition}, a system in thermal equilibrium distributes its energy equally among all its energetic components. And indeed, as it is demonstrated in \cite{thermalizationCX}, after the thermalization time $t_{eq}$, the expected current- and osmotic- kinetic energies (and expected quantum potential), all become equal to each other and each equal to half of the standard kinetic energy expectation. This implies that the times $t>t_{eq}$ (after thermalization), are characterised by the condition $\langle \hat{P}^2_j/(2m)\rangle\simeq2\langle K_{B,j}\rangle\simeq2\langle K_{O,j}\rangle\simeq2\langle Q_{j}\rangle$. Figures \ref{fig1}.(c), (f) and (i) demonstrate that the ``hidden"-variables $\langle K_{B,j}\rangle$ and $\langle Q_{j}\rangle$ (or $\langle K_{O,j}\rangle$) converge to equal values after $t_{eq}$, effectively revealing the onset of thermalization. Importantly, as illustrated in the figure, this holds true even when Hermitian position or velocity expectations, as well as the expectation of kinetic energy itself, are insensitive to the thermalization process. 

Anyone familiar with any of the aforementioned theories, {\em regardless of their ontological preferences,} could potentially arrive at this conclusion. To us, this success of certain functions of weak values in characterizing a seemingly pathological quantum system clearly illustrates the {\em predictive} power and physical interest of weak values. Moreover, this case study also shows the {\em heuristic} potential inherent in Bohmian-like theories that can serve as a guide to find the relevant functions. \vspace{-0.2cm}

\section{Final Remarks}

In this chapter we have shown how weak values are very useful predictive and characterization tools, regardless of ontological disputes, and how certain hidden variables theories provide heuristic frameworks that guide us towards physically informative weak values. To illustrate these conclusions, as a case study, it is shown how the equlibration time in certain scenarios is better determined from functions of weak values borrowed from the Bohmian machinery, rather than from standards operators.  To properly acknowledge the merits of our work, we emphasize that all results shown in section \ref{casestudy} in terms of (functions of) \emph{Bohmian-inspired} weak values are not merely theoretical embellishments but also outcomes directly accessible in the laboratory following the protocol outlined in subsection \ref{emp1}. This connection between theoretical and experimental \emph{Bohmian-inspired} weak values opens up new and unexplored possibilities for characterizing quantum systems, with quantum thermalization discussed in this chapter serving as just the first example.\vspace{-0.05cm}

Let us finish with a final, more speculative remark. It is relevant to question how one should approach elements of a theory that prove to be useful in enhancing its predictive power. Should they be regarded merely as useful fictions or as insights into the inner workings of reality? Consider, for example, the Copernican system with its heliocentric hypothesis. As it is well-known, in the preface of \textit{De Revolutionibus} written by Andreas Osiander, this hypothesis was deemed a mere useful fiction that would allow astronomers to perform certain calculations regarding planetary movements with more ease. However, the heliocentric hypothesis ended up considered not merely useful but true. In 1808, John Dalton used the “imaginative” and “hidden” notion that matter consists of atoms to explain fixed ratios in chemical combinations. For several decades, the mainstream resisted the ontological implications of this calculation method (“who [has] ever seen a gas molecule or an atom?”, was the objection by Marcelin Berthelot still in 1877 \citep{quantumreality}). Yet, as we know, the story concluded with the atomic hypothesis being accepted as true. Although the atomic hypothesis has undergone significant refinement, and our current conception of what an atom is substantially differs from what Dalton believed, we still acknowledge the validity of his insight. Now, we ponder whether a similar narrative could unfold with weak values and theories like Bohmian mechanics, not as the ultimate truths (we are fully aware that we are dealing with non-relativistic quantum phenomena), but as indicative of the right path to follow. May we be experiencing a “déjà vu” in this regard? Only time will tell…

\begingroup
\renewcommand{\addcontentsline}[3]{}

\def\bibsection{\section*{V.$\quad$REFERENCES\vspace{0.3cm}}} 

{\footnotesize
\bibliography{Bibliography.bib} 
}
\endgroup

\addcontentsline{toc}{section}{$\ $V. References}

\appendix

\begingroup
\renewcommand{\addcontentsline}[3]{}
\section{On the weak value $\w_{e,i}$}
\endgroup
\addcontentsline{toc}{section}{VI. Appendices}
\addcontentsline{toc}{subsection}{A. On the weak value $\w_{e,i}$}

When considering the experimental protocol for the weak value $\w_{e,r}$, we have assumed, first, that there is a weak interaction between the object system and the ancilla. As a result of this interaction, the latter is (slightly) displaced as a function of the momentum of the object system. But, of course, it also acquires some momentum itself. In the protocol for $\w_{e,r}$, the displacement of the ancilla is measured right afterwards. If instead of the position, one measures its momentum (let us denote $\s^{(k)}$ for the outcome of $k$-th repetition), while keeping the rest of the protocol unchanged, the weak value $\w_{e,i}$ is obtained. Thus, by replacing $\m$ with $\s$ in the probabilities after \eqref{pcond}, the following asymptotic definition of $\w_{e,i}$ results,\vspace{-0.2cm}
\begin{equation}\label{wvi}
    \w_{e,i}(x,t,\tau):=\lim_{M\rightarrow \infty}\frac{\sum_{k\in\sigma} \s^{(k)}}{M_x}=\frac{\int d\s\: \s\:\p(\s,x\,|\,t,\tau)}{\int d\s\:\p(\s,x\,|\,t,\tau)}.
\end{equation}
Next, in order to derive the formal expression, we note that the difference with the protocol for $\w_{e,r}$ occurs after the object-ancilla entanglement, when the ancilla is measured. In this case, the momentum of the ancilla (not its displacement) is strongly coupled with the macroscopic pointer. By rewriting the composite's entangled state \eqref{onepm} as\vspace{-0.2cm}
\begin{equation}
    |\Psi_{pre}\rangle = \iint \psi(p_s)\tilde{f}(p_a) e^{-i\gamma Tp_sp_a/\hbar} |p_s\rangle |p_a\rangle dp_s dp_a,\vspace{-0.2cm}
\end{equation}
using the Fourier transform of the ancilla's wavepacket, $\tilde{f}(p_a):=\frac{1}{\sqrt{2\pi\hbar}}\int e^{-ip_ay/\hbar}f(y)dy$, the read-out of $\s$ for the ancilla's momentum collapses the system into the perturbed state 
$|\tilde{\psi}_{\s}(t)\rangle = \hat I _s\otimes \langle{\s}|\Psi_{pre}\rangle_{sa}=\int \psi(p_s)\tilde{f}(\s) e^{-i\gamma Tp_s\s/\hbar} |p_s\rangle dp_s .$ The free temporal time evolution, denoted by $\hat{U}_s(\tau)$, leaves the object's state as $\hat{U}_s(\tau)|\Tilde{\psi}_{\s}\rangle$.

For a sufficiently small object-ancilla interaction, $\gamma T$ is such that, for all $p_s$, one can truncate $e^{-i\gamma T p_sp_a/\hbar}=1-i\gamma T p_sp_a/\hbar + O((\gamma T)^2)$ at first order. Introducing this expansion in the probability density $\p(\s,x\,|\,t,\tau)=|\langle x|\hat{U}_s(\tau)|\Tilde{\psi}_{\s}\rangle|^2$, obtained by Born's rule, one gets
\begin{equation}
    \p(\s,x\,|\,t,\tau)=|\tilde{f}(\s)|^2\qty{ |\langle x|\hat U_s(\tau)|\psi(t)\rangle|^2-2\frac{\gamma T \s}{\hbar}\Re\qty[i\langle \psi(t)|\hat U^\dagger_s(\tau)|x\rangle\langle x|\hat U_s(\tau) \hat{P}_s|\psi(t)\rangle]}+O((\gamma T)^2).
\end{equation}
Let us assume that the expected momentum of the ancilla's rest state is calibrated at zero, $\int|\tilde{f}(p_a)|^2p_a\: dp_a=0$. It will have a non-zero variance $\Delta^2\propto 1/\sigma^2$, since the position-space wavepacket $f(y)$ has a finite spread $\sigma$. Then,
\begin{equation}
    \int \s \: \p\:(\s,x\,|\,t,\tau) d\s = -\frac{2 \gamma T\Delta^2}{\hbar}|\langle x|\hat U_s(\tau)|\psi\rangle|^2\Re\Bigg[i\frac{\langle x|\hat U_s(\tau)\hat{P}_s |\psi(t)\rangle}{\langle x|\hat{U}(\tau)|\psi(t)\rangle}\Bigg]+O((\gamma T)^2)
\end{equation}
and $    \int \: \p(\s,x\,|\,t,\tau) d\s =|\langle x|\hat U_s(\tau)|\psi\rangle|^2+O((\gamma T)^2)$.
With all, by choosing the ancilla wavepacket to be a minimal uncertainty state, such that $\Delta^2 = \hbar/2$, one can obtain, as anticipated, that to leading order,
\begin{equation}
    \w_{e,i}(x,t,\tau)=\frac{\int d\s\: \s\:\p(\s,x\,|\,t,\tau)}{\int d\s\:\p(\s,x\,|\,t,\tau)}\propto \Im\Bigg[\frac{\langle x|\hat U_s(\tau)\hat{P}_s |\psi(t)\rangle}{\langle x|\hat{U}(\tau)|\psi(t)\rangle}\Bigg].
\end{equation}

\begingroup
\renewcommand{\addcontentsline}[3]{}
\section{Generalization to many-body weak values}
\endgroup
\addcontentsline{toc}{subsection}{B. Generalization to many-body weak values\vspace{0.7cm}}

In this section, we generalize weak values to many-body quantum systems with an arbitrary number $N$ of degrees of freedom. 
We will call them just $N$ {\em particles}, considering without loss of generality a physical space of one dimension. 

Let us start from the $N=2$ case. We define the experimental weak value $\w_{e,1}(x_1,x_2,t)$ as the asymptotic average across repeated preparations of the weak measurement of the momentum of the first particle (of operator $\hat p_1$), conditioned on the particles being found at $x_1$ and $x_2$ respectively. We define $\w_{e,2}(x_1,x_2,t)$ by changing the weak measurement of particle $1$ to that of particle $2$. The key point here is that we need to identify which is the particle $1$ and which the $2$, both for the momentum measurement and the subsequent simultaneous position measurements. 

In general, for $N$ {\em distinguishable} particles, such an identification is not fundamentally problematic and all the empirical, formal and ontological conclusions from section II follow straightforwardly. As such, the formal many-body weak values for $N$ distinguishable particles can be defined as
\begin{equation}
    \w_{f,j}(\vec{x},t):=\frac{\langle \vec{x}|\hat P_j|  \psi(t) \rangle}{\langle \vec{x}|  \psi(t) \rangle},
    \label{modal3}
\end{equation}
where the subindex $j\in\{1,...,N\}$, just specifies the degree of freedom that is being measured. Note that we have replaced the scalar $x$ in physical space by the configuration-space vector $\vec{x}=\{x_1,x_2,...,x_N\}$. Its implications are clear in Bohmian mechanics, where $Re[\w_{f,j}(\vec{x},t)]$ is the momentum of the $j$-th particle {\em when} the whole system has a configuration $\vec{x}$. This means (now for any theory) that the velocity information of one particle depends on all the rest of particles. This is a manifestation of quantum entanglement and non-locality. Beyond this dependence on configuration-space, there is no relevant novelty in dealing with many-body weak values for distinguishable particles. 

For {\em indistinguishable} particles, the former conclusions need to be revisited with some care. Formally, we could still compute the same weak values of \eqref{modal3} from the wavefunction.\footnote{This is why in Bohmian-like theories indistinguishable particle trajectories are still mathematically distinguishable and well-defined.} However, empirically, when we measure the momentum or position of one of the indistinguishable particles (weakly or not), it will be impossible to know which of the particles we are referring to. Due to this inherent ambiguity, an alternative weak value protocol that eliminates the need for particle tagging must be proposed. In order to achieve this, it is important to recognize that indistinguishable particles render configuration space empirically inaccessible. Instead, we can devise a protocol in which a particle-agnostic weak measurement of momentum in a particular direction of {\em physical space} $x$, is followed by the determination of the position of {\em one} of the particles. As shown in \cite{manybody1}, the resulting formal weak value $\tilde P_f$, is roughly the average of the weak values that would correspond to each particle if they were {\em distinguishable}\vspace{-0.15cm}
\begin{equation}
\tilde P_f(x,t) = \frac{1}{N^2} \sum_{j=1}^{N} \sum_{k=1}^{N} P_{f,j,k}(x,t),
\label{wva}
\end{equation}
where $ P_{f,j,k}$ is the conditional marginalization of the experimentally {\em inaccessible} weak value $\w_{f,j}$, 
\begin{equation}
P_{f,j,k}(x,t) = \left[\frac{ \int d\vec x_k  \; \w_{f,j}(\vec{x},t) |\Psi(\vec{x},t)|^2}{\p(x_k,t)} \right]_{x_k=x}
\label{wvp}
\end{equation}
and $\vec{x}_k:=\{x_1,..,x_{k-1},x_{k+1},...,x_N\}$. Note that $\p(x_k,t):=\int d\vec x_k   |\Psi(\vec{x},t)|^2$
is the probability of finding one particle at $x$ no matter where the others are.\footnote{Due to symmetry considerations, the non-integrated index $k$ is not relevant in the definition of $\p(x_k,t)$.}

It is worth noticing that these new weak values still are understandable as local expectation values, so that the (average\footnote{Standard expectations for individual particles are also hidden. Only the average standard expectation is experimentally accessible.}) standard expectation can still be retrieved from them by averaging,\footnote{For indistinguishable particles, any two particles, say $j$ and $k$, have the same expectation value $\langle \hat P \rangle(t)=\langle \hat P_j \rangle(t)=\langle \hat P_k \rangle(t)$.}
\begin{equation}
 \langle \hat P \rangle(t):=\frac{1}{N}\sum_{j=1}^{N} \langle \hat P_j \rangle(t)=\int dx \tilde P_f(x,t) \p(x,t).
\label{wvaid}\vspace{-0.2cm}
\end{equation}
 Importantly, notice that $P_{f,j,k}(x,t)$ in \eqref{wvp} already contains $N-1$ of the $N$ spatial integrals of  $\langle \hat P_j \rangle(t)$ in the right hand side of \eqref{wvaid}. Thus, for a large $N$ (characteristic of many-body systems) we obtain $\tilde P_f(x,t) \approx \langle \hat P_j \rangle(t)$. This implies that the weak values obtained for indistinguishable particles provide almost the same information as the standard expectation values. Initially, this may suggest that they are not as practical as weak values obtained for distinguishable particles. However, in realistic many-body quantum systems, individual particle information tends to be uninteresting and, instead, the focus is on holistic degrees of freedom such as the physical-space center of mass. Remarkably, the center of mass is always experimentally distinguishable, and from a formal perspective, the many-body Hamiltonian can typically be separated into two parts: one for the center of mass and another for all the relative coordinates. Consequently, weak values obtained for the center of mass retain the most informative aspects of distinguishable particle weak values (but now in physical space).


\end{document}